    \documentclass[12pt]{iopart}
	\usepackage{amssymb}
	\usepackage{amsthm} 
	\usepackage{bm}
	\usepackage{enumerate}
	\usepackage{graphicx}
	\usepackage[caption=false]{subfig}
	\usepackage[colorlinks,citecolor=red,linkcolor=blue,urlcolor=blue]{hyperref}

	\newcommand{\abs}[1]{\left| #1 \right|} 
	\newcommand{\avg}[1]{\left< #1 \right>} 
	
\begin{document}
	
	\title[Phase Diagram of the SS-KLM with Classical Localized Spins]{Phase Diagram of the Shastry-Sutherland Kondo Lattice Model with Classical Localized Spins: A Variational Calculation Study}
	
	\author{Munir Shahzad and Pinaki~Sengupta}
	
	\address{School of Physical and Mathematical Sciences, Nanyang Technological University, 21 Nanyang Link, Singapore 637371}
	
	\ead{munir001@e.ntu.edu.sg, psengupta@ntu.edu.sg}
	\vspace{10pt}
	
	\begin{abstract}
	We study the Shastry-Sutherland Kondo lattice model with additional Dzyaloshinskii-Moriya~(DM) interactions, exploring the 
possible magnetic phases in its multi-dimensional parameter space. Treating the local moments as classical spins and using a variational ansatz, we identify the parameter ranges over which various common magnetic orderings are potentially stabilized. Our results reveal that the competing interactions result in a heightened susceptibility towards a wide range of spin configurations including longitudinal ferromagnetic and antiferromagnetic order, coplanar flux configurations and most interestingly, multiple non-coplanar configurations including a novel \textit{canted-Flux} state as the different Hamiltonian parameters like electron density, interaction strengths and degree of frustration are varied. The non-coplanar and non-collinear magnetic ordering of localized spins behave like emergent electromagnetic fields and drive  unusual transport and electronic phenomena. 
	\end{abstract}
	%
	\vspace{2pc}
	%
	
	\section{Introduction}\label{sec:intro}

	Spin-charge coupled systems on geometrically frustrated lattices have turned out to be a promising avenue to look for the novel and exotic phases. One intriguing property of frustrated systems is their degenerate ground state -- which makes them more susceptible and sensitive to small perturbations such as thermal  fluctuations, disorder, anisotropy or long-range interactions\cite{lacroix-2011,diep-2013,balents-2010,moessner-2006}. In real materials, often the coupling of spins to other degrees of freedom partially lifts the ground state degeneracy. For example, in frustrated spin-orbital systems the magnetic phase transition is mediated by orbital order\cite{pen-1997,tchernyshyov-2004,garlea-2008,gwchern-2010,tokura-2000}. In case of hybrid systems itinerant electrons interacting with localized spins arranged on geometrical frustrated lattices have given rise to unconventional phases and physical phenomena\cite{taguchi-2001,chern-2011,akagi-2015,reja-2016,ishizuka-2013}. Prominent among these new states are the chiral ordered states with non-zero scalar spin chirality\cite{kumar-2010,hayami-2014,batista-2015}. The chiral nature of these states breaks both parity and time-reversal symmetries. Such non-coplanar spin orderings induce novel quantum phases and phenomena (that are not observed for their coplanar or collinear counterparts) such as the geometric or topological Hall effect (THE) where there is a Hall conductivity even in the absence of an external applied magnetic field\cite{kenya-2000,shindou-2001,martin-2008,chern-2012}. The itinerant electrons acquire an extra Berry phase when its spin follows the spatially varying magnetization present in these spin textures -- equivalent to an orbital coupling to a magnetic field. THE  has already been observed in ferromagnetic pyrocholre lattices {Pr$_2$Ir$_2$O$_7$} and {Nd$_2$Mo$_2$O$_7$}\cite{yoshii-2000,machida-2007,udagawa-2013}. Chiral spin orders have been studied theoretically in the context of Kondo Lattice Model (KLM) on frustrated lattices such as triangular\cite{yasu-2010,kipton-2013,rahmani-2013}, kagome\cite{kenya-2000,kipton-2014,chern-2014,ghosh-2016}, pyrocholre\cite{chern-2010}, face-centered cubic lattice\cite{shindou-2001} and checkerboard lattice\cite{venderbos-2012}. Our plan is to extend this study to the geometrically frustrated Shastry-Sutherland Lattice (SSL). 
    This is not simply of academic interest. There exists a complete family of rare earth tetraborides, {RB$_4$}, (R=Tm, Er, Tb, Dy, Ho) where there is a strong coupling between the itinerant electrons arising (predominantly) from the unfilled $5d$ orbitals of the R$^{3+}$ ions and localized moments due to unfilled $4f$ orbitals of the same\cite{siemensmeyer-2008,keola-2015,sai-2016,suzuki-2010,matas-2010,suzuki-2009,michimura-2009}. The R$^{3+}$ ions in these magnets are arranged in an SSL geometry, making the Shastry-Sutherland Kondo Lattice Model (SS-KLM) the ideal framework to understand the magnetic and electronic properties of this family of metallic quantum magnets\cite{shin-2017}. We have chosen a complete general set of DM interactions and identified the minimal set of DM interactions necessary for the present model to stabilize the chiral spin states. 
Owing to the large number of parameters involved, the SS-KLM is expected to have an extremely rich phase diagram with multiple competing ground state phases stabilized in different parameter regimes. Our goal in this paper is to identify ranges of parameters that favor non-coplanar spin orderings. With that goal, and following previous studies on similar models\cite{akagi-2010,akagi-2011,akagi-2013}, we explore the ground state phase diagram of this model with variational calculations and find out the parameters space where chiral spin phases are energetically favored.  Using a variational ansatz, we compare the energies for a variety of commonly observed ground state phases and determine the state with minimum energy. While such an approach may not always yield the true ground state, it is an effective way of identifying parameter regimes with maximum susceptibility towards the desired ground states. Our results suggest that chiral spin arrangements are stabilized over a large range of parameters when we include the DM interaction. We have obtained a new non-coplanar \textit{canted-Flux} state which is stabilized not only for insulating phases but also for metallic phases. 


    \section{Model}\label{sec:model}

\begin{figure}[htb]
	\centering
	\includegraphics[width=0.5\linewidth]{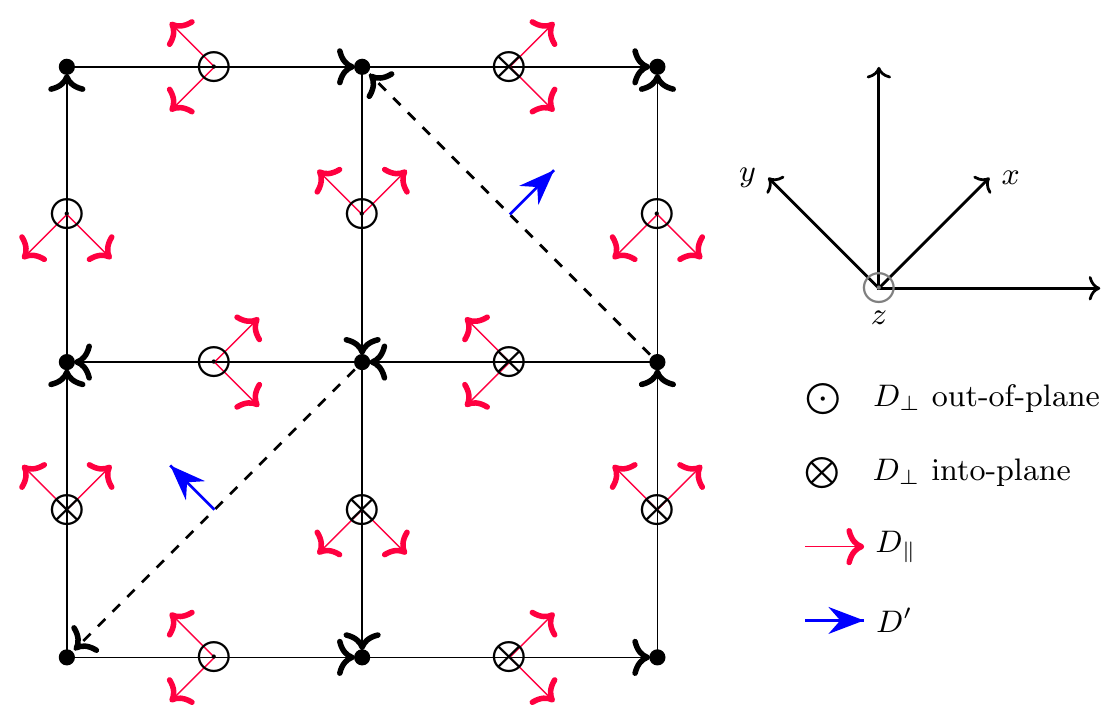}
	\caption{DM interaction defined on the unit cell of SSL where direction of arrow from site $i$ to $j$ indicates the direction of cross product $\mathbf{S}_i\times \mathbf{S}_j$. The red arrows represent the in-plane components of $\mathbf{D}$ while $\bigotimes$ and $\bigodot$ represent into and out-of-plane components of $\mathbf{D}$. Blue arrows indicate the components of $\mathbf{D}'$ on the diagonal bonds. The directions of $x$, $y$ and $z$ axis are also mentioned.}
	\label{fig:SSL-dm04}
	\end{figure}

	We study the Kondo Lattice model with additional DM and Heisenberg exchange interactions on the geometrically frustrated SSL with classical localized spins. The total Hamiltonian can be written as,
	
	\begin{eqnarray}
	\label{equ:ham04}
	\mathcal{H}= & \underbrace{-\sum_{\avg{i,j},\sigma}t_{ij}(c_{i\sigma}^\dagger c_{j\sigma}+\mathit{h.c.})-J_K \sum_i \mathbf{S}_i\cdot \mathbf{s}_i}_{\mathcal{H}_e} \nonumber
	\\
	& +\underbrace{ \sum_{\avg{i,j}}J_{ij}\mathbf{S}_i \cdot \mathbf{S}_j + \sum_{\avg{i,j}}\mathbf{D}_{ij}. (\mathbf{S}_i\times \mathbf{S}_j)}_{\mathcal{H}_c}
	\end{eqnarray} 
	The first two terms constitute the electronic part of the Hamiltonian $\mathcal{H}_e$ where we have a tight binding term for itinerant electrons and an on-site Kondo-like interaction between the spin of these electrons and localized moments. $\avg{i,j}$ denotes the bonds on the SSL, where nearest neighbors (NN) bonds are axial bonds and next nearest neighbors (NNN) are alternate diagonal bonds and $t_{ij}$ represents the transfer integral for itinerant electrons hopping on these SSL bonds.  In the present study, we ignore any interaction between the itinerant electrons.~\cite{liu-2014,chen-2012} The localized spins $\mathbf{S}_i$ are treated as classical vectors of unit length. The sign of $J_K$ then becomes irrelevant in the current model as the eigenstates that correspond to different sign of $J_K$ can be related by a global gauge transformation\cite{pekker-2005,martin-2008}. The last two terms represent the interaction between localized spins. In this part of the Hamiltonian $\mathcal{H}_c$, the first term represents anti-ferromagnetic exchange interaction between the localized spins. The last term is the DM interaction and more detail about this interaction on NN and NNN bonds is presented in the caption of Fig.~\ref{fig:SSL-dm04}. Hereafter the primed parameters represent the interactions on diagonal bonds while the unprimed are for axial bonds. We choose $t=1$ as the energy unit.
	
	\begin{figure}[htb]
	\centering
	\includegraphics[width=0.99\linewidth]{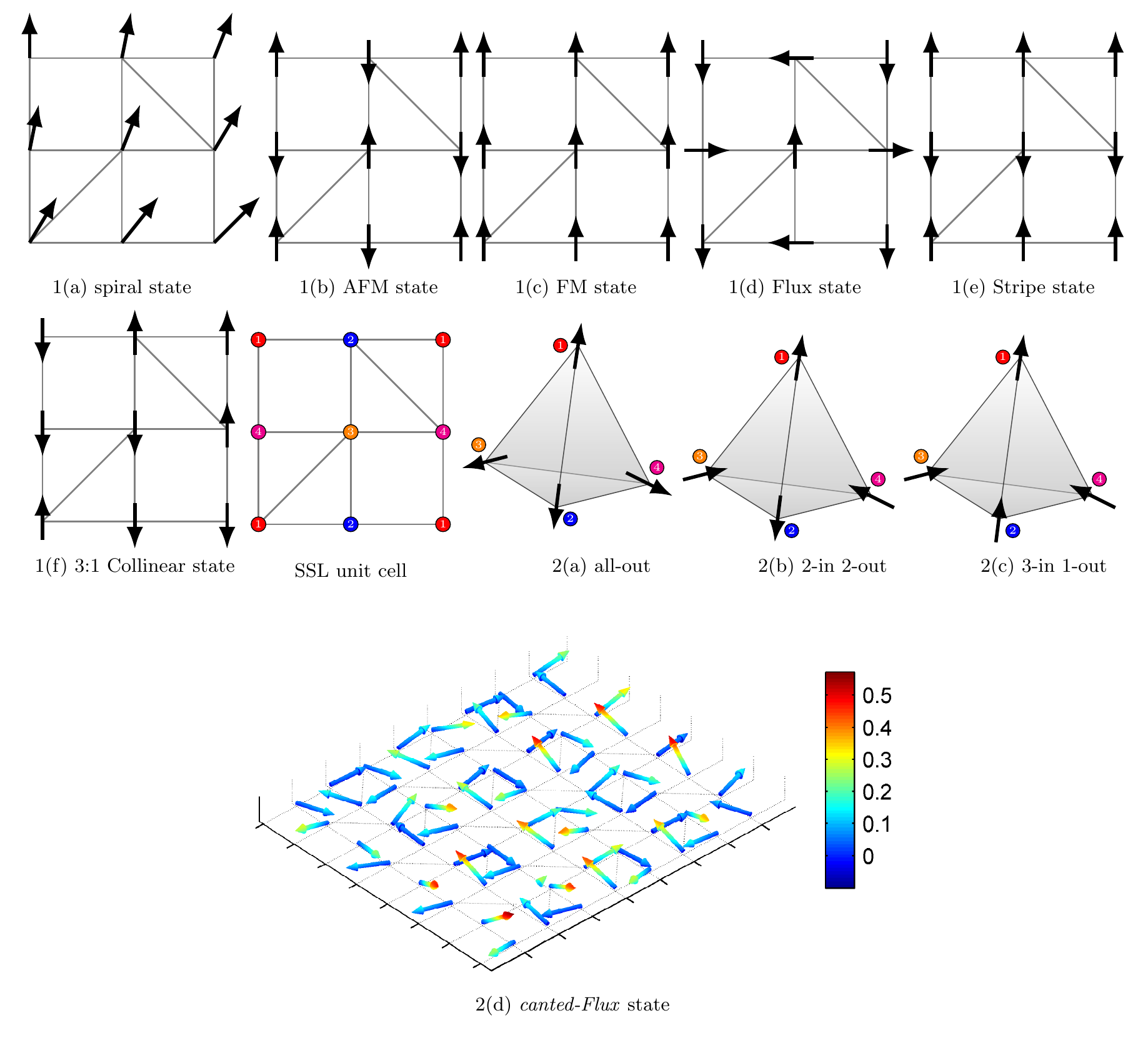}
	\caption{ Different ordered configurations used in the variational calculations 1(a) Spiral state 1(b) AFM state 1(c) FM state 1(d) coplanar Flux state 1(e) stripe state 1(f) 3:1 collinear state 2(a) all-out state 2(b) 2in 2-out 2(c) 3-in 1-out state 2(d) \textit{canted-Flux} state (here this ground state is shown on $8\times8$ lattice). From 1(a)-1(f) are coplanar arrangements of localized spins while 2(a)-2(d) represent the non-coplanar configurations. See text for more details.}
	\label{fig:all-states}
	\end{figure}
	
	\section{Method}\label{sec:methods}
	
	We study the phase diagram of the Hamiltonian~(\ref{equ:ham04}) as a function of itinerant electron number density $n_e=\frac{1}{2N}\sum_{i\sigma}\avg{c_{i\sigma}^\dagger c_{i\sigma}}$ and Kondo interaction $J_K$ with the help of variational ansatz at $T=0$. We estimate the total energy of the system for different configurations of localized moments and then identify the most energetically favorable state having minimum energy between all of them. By replacing the itinerant electron spin as $\mathbf{s}_i
	=c_{i,\alpha}^\dagger\bm{\sigma}_{\alpha\beta}c_{i,\beta}$ the electronic part of the Hamiltonian in Eq.~(\ref{equ:ham04}) becomes quadratic in fermionic operators, where $\bm{\sigma}_{\alpha \beta}$ are vector elements of usual Pauli matrices. This transformation also helps us to write this part as $2N \times 2N$ matrix for a particular configuration of the localized spins using single particle basis, where $N$ is the number of sites. The electronic energy for a specific ordering of localized spins can then be calculated by diagonalizing this matrix. The energy that corresponds to the localized spin part of the Hamiltonian ${\mathcal{H}_c}$ is obvious to calculate once we have specified the ordering of these classical spins. The total average energy then can be written as,
	
	\begin{eqnarray}
	\label{equ:enr}
	\avg{E}= \sum_i f(\epsilon_i)\epsilon_i+ \sum_{\avg{i,j}}J_{ij}\mathbf{S}_i \cdot \mathbf{S}_j + \sum_{\avg{i,j}}\mathbf{D}_{ij}. (\mathbf{S}_i\times \mathbf{S}_j)
	\end{eqnarray}
	where $f(\epsilon)$ is the Fermi distribution function and at $T=0$ this will be non-zero only when $\epsilon\leq\epsilon_f$.
	
	In the calculations, we considered $10$ different ordered configurations of localized spins including coplanar and non-coplanar states. These states are shown in Fig.~\ref{fig:all-states}, where spiral, anti-ferromagnetic (AFM), ferromagnetic (FM), Flux, stripe and 3:1 collinear are coplanar ordered states while all-out, 2-in 2-out,3-in 1-out and \textit{canted-Flux} are non-coplanar states. The spiral state chosen in the study is the ground state for spin only Shastry-Sutherland model for the values of Heisenberg interaction considered here\cite{shastry-1981,moliner-2009,grechnev-2013}. For this state the angle difference between nearest neighbor spins is $\phi=\pi \pm \arccos(\frac{J}{J'})$. We have also considered a \textit{canted-Flux} state that may arise due to the interplay between the Heisenberg and the DM interactions. For the phase diagram as a function of $n_e$, we need to look the relationship between $n_e$ and chemical potential $\mu$. Generally, the phase transition between different magnetic states is discontinuous accompanied by a jump of $n_e$. We have obtained the ground state by comparing energies of all these ordered states by varying $\mu$. We have also calculated the electron number density $n_e$ as a function of $\mu$. At the end, the phase diagram is obtained by mapping ground state energy versus $\mu$ curve on $n_e$ versus $\mu$ curve. Normally at $T=0$, the phase transitions between different magnetic states are of first order and accompanied by a phase separation (PS) region. This PS region is characterized by a jump of $n_e$ as in this region the system is not stable and can not have a fixed number density\cite{akagi-2010,akagi-2011,akagi-2013}.
	
    \section{Results}\label{sec:results}
	
\begin{figure}[htb]
	\centering
	\subfloat[]{\label{fig:nodm}{\includegraphics[width=0.8\linewidth]{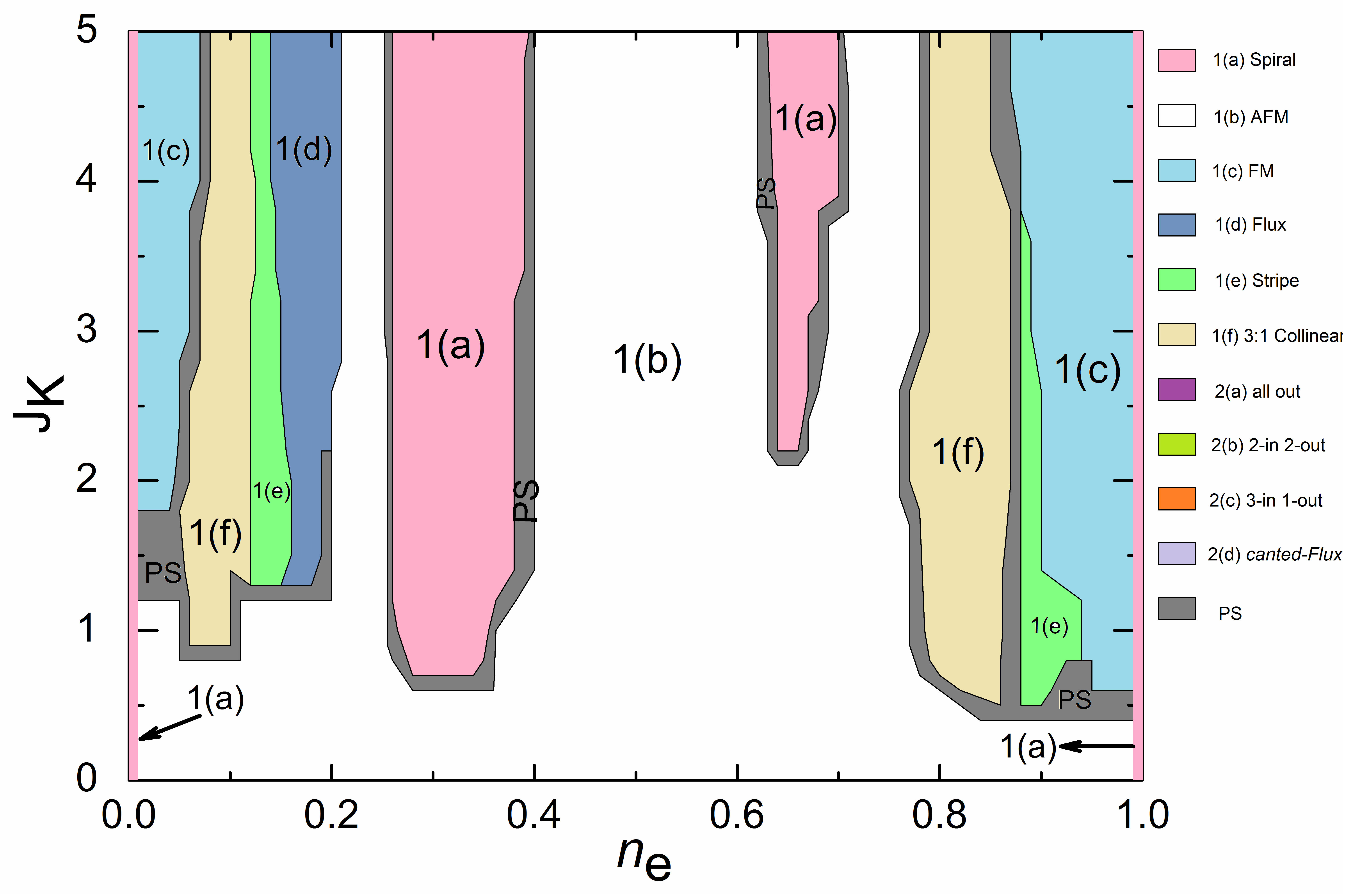}}}
	
	\subfloat[]{\label{fig:withdm}{\includegraphics[width=0.8\linewidth]{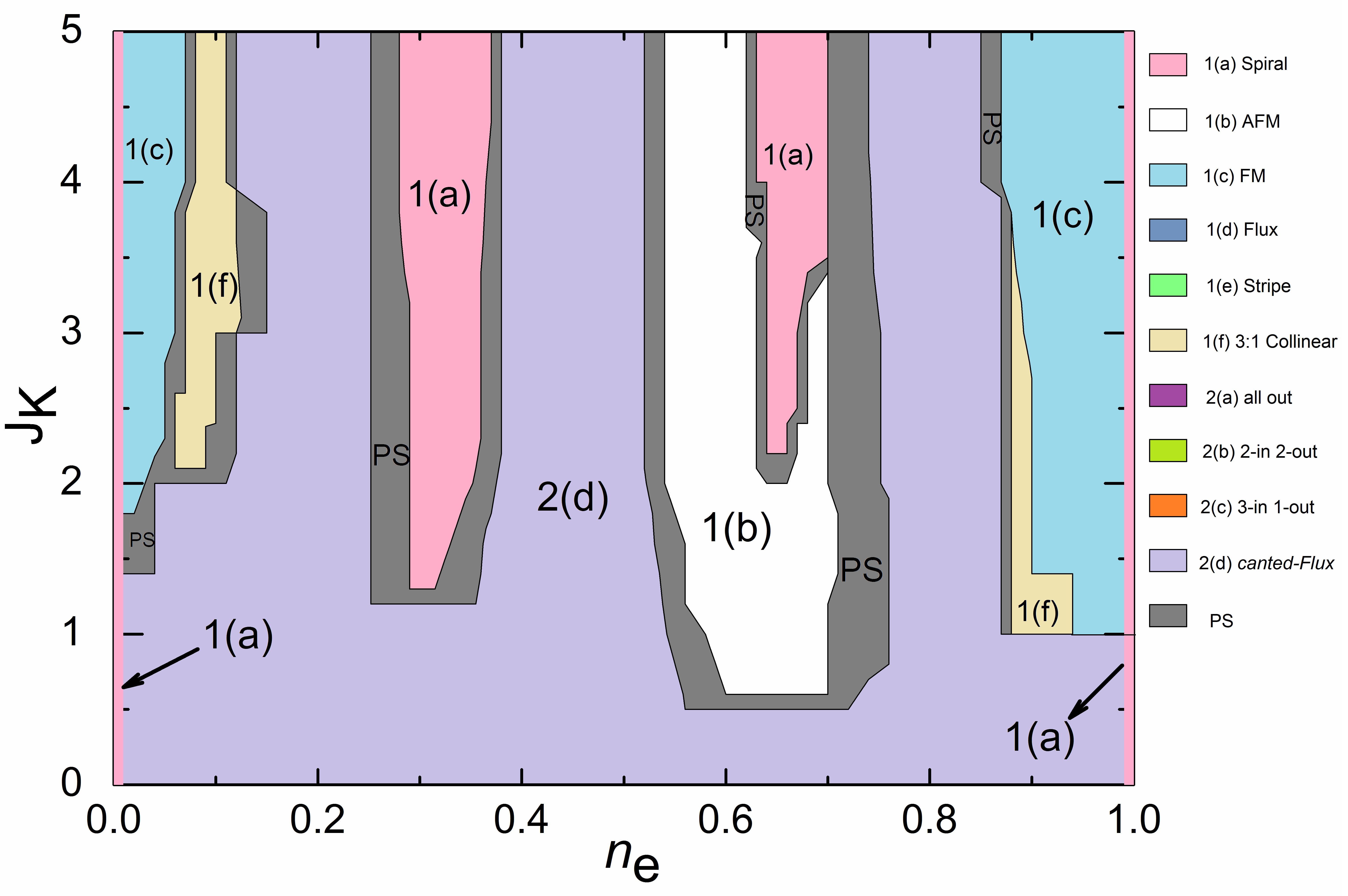}}}
	\caption{The phase diagram at $t=t'=1.0$, $J=0.1$, $J'=0.12$ (a) $\abs{\mathbf{D}}=\abs{\mathbf{D}'}=0$ (b) $D_\parallel=-0.1$, $D_{\perp}=0.1$ and $D'=-0.12$. The calculations are done on $80 \times 80$ lattice and the regions marked with 1(a)-1(f) and 2(a)-2(d) indicate the portions where ordered states in Fig.~\ref{fig:all-states} are stabilized. PS describes a phase separation region. The magnetic ground state is a spiral phase for all values of $J_K$ at $n_e=0$ and $1$.}
		
	\end{figure}

	We have performed the above mentioned variational calculations on finite sized lattices 
while varying the parameters $n_e$, $J_K$, for several representative sets of strengths of 
Heisenberg exchange interaction and DM vector.  The results presented were obtained for a 
$80\times80$ lattice, and they were verified on other sizes at a few values of the parameters. 
The chemical potential $\mu$ and electron density $n_e$ are divided into $O(10^4)$ 
points ($12,800$ to be exact). Since all the states have a periodicity of at most $2$ unit cells,  
the results from the $80\times 80$ lattices were found to be adequate and no significant finite 
size effects were observed. In all cases, the local moments were found to be in a spiral 
configuration at $n_e = 0$ and 1. In these limits (which correspond to empty and completely 
filled electrons band), the Kondo coupling, $J_K$, has no meaning and the model reduces to 
classical moments with competing interactions on the SSL. This has been studied in 
Ref.~\cite{moliner-2009} and our results are consistent with those reported therein.
However, at infinitesimally small deviations from these limits, there is a discontinuous
transition from the spiral phase. The nature of the magnetic ordering at intermediate to
large values of $J_K$ strongly depends on the filling factor, whereas at small $J_K$, the 
spin configurations are independent of the filling factor. We start our discussion 
with Fig.~\ref{fig:nodm} that shows the phase diagram without DM interaction for $t'=1.0$, 
$J=0.1$, and $J'=0.12$ as a function of $n_e$ and $J_K$. At very small and high value of 
electron number density FM metallic ground state is stabilized and this region becomes wider 
as $J_K$ increases. This part of the phase diagram appears 
as a result of double exchange mechanism\cite{zener-1951,anderson-1955}. The fermionic kinetic energy (K.E.) stabilizes the 
FM ordering of the localized spins as there is large K.E. gain if the spins on two sites are 
parallel. However, at half filling this argument is not valid as the lower bands are completely 
filled and we need energy of the order of $J_K$ to cause the hopping. The second order 
perturbation theory in $t/J_K$ yields an effective Heisenberg interaction with the coupling 
constant $t^2/J_K$ which favors the AFM ordering of the localized spins. So for the current value 
of parameters, an insulating AFM ground state is realized around half filling. At intermediate 
filling factors there is a competition between these two effects that leads to spiral phase, the 
states that are mixture of FM and AFM states and PS regions. Accordingly, the phase diagram 
[see Fig.~\ref{fig:nodm}] in the paramter space of $n_e$ vs. $J_K$ features other coplanar 
states such as spiral, flux, stripe and 3:1 collinear states. Recent study on double exchange 
model~\cite{maria-2017} shows that FM is unstable against a non-collinear spiral phase at 
commensurate filling -- though in our model we chose a different spiral state but occurrence of 
such phase in the phase diagram reinforces their point. The transitions between two ordered 
states is accompanied by PS regions in the phase diagram and in these parts the ground state 
has a volume which is mixture of the ordered states\cite{dagotto-2003,kagan-1999}. As 
mentioned in the introduction part we are interested in the phases with non-zero chirality 
which are absent in the Fig.~\ref{fig:nodm}.

	What happens to phase diagram when we \textit{switch on} the DM interaction? The results are shown in Fig.~\ref{fig:withdm}. Clearly, DM interaction stabilizes the states that are non-coplanar in nature and that is why the regions where chiral states are realized becomes wider in size. In fact what we found out is that the \textit{canted-Flux} state is the one that now covers most part of the phase diagram. This state is a novel chiral phase that has not been observed earlier. This state driven by DM interaction replaces the AFM at $n_e=1/2$, $1/4$ and $3/4$. We focus on this state at half filling where the system is in insulating phase. We found out that this phase is stabilized even at finite temperature.  Details on dynamically stabilizing this ground state using MC updates of some initial random configurations will be presented elsewhere\cite{munir-2017}. Moreover, with the increase of the $J_K$ interaction the portions with \textit{canted-Flux} phase shrink and at very high value of $J_K$ around $n_e\simeq3/4$ what we get is a PS. FM phase appearing at small and high values of number density is robust against this DM interaction. Among other coplanar states that still survive in the presence of DM interaction are spiral, 3:1 collinear and some portions of AFM phase.
	
	\begin{figure}[htb]
	\centering
	\subfloat[]{\label{fig:nodm2}{\includegraphics[width=0.8\linewidth]{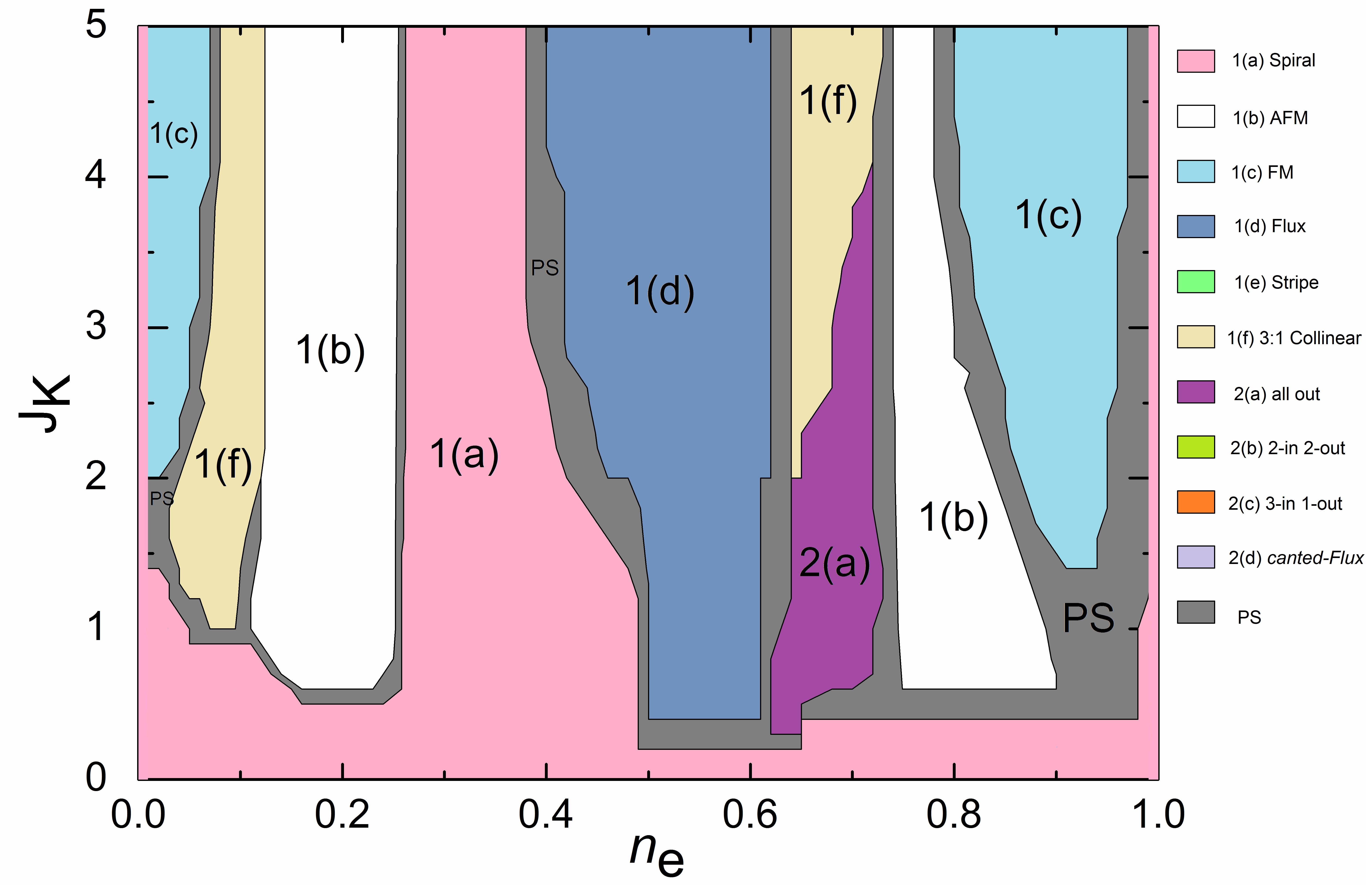}}}
		
	\subfloat[]{\label{fig:withdm2}{\includegraphics[width=0.8\linewidth]{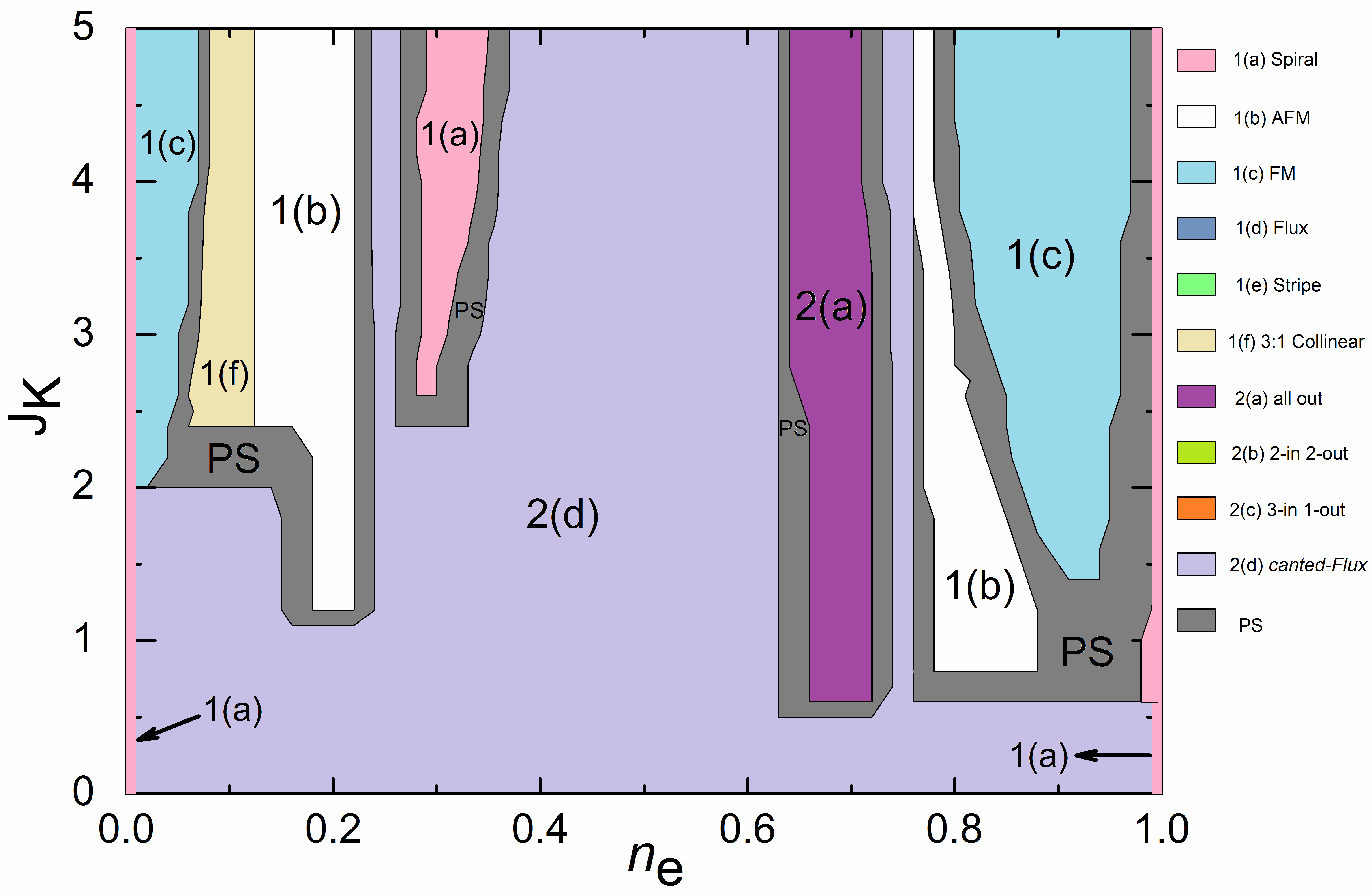}}}
	\caption{The Phase diagram at $t=1.0$, $t'=2.2$, $J=0.1$, $J'=0.22$ (a) $\abs{\mathbf{D}}=\abs{\mathbf{D}'}=0$ (b) $D_\parallel=-0.1$, $D_{\perp}=0.15$ and $D'=-0.22$. The calculation are done on for $80 \times 80$ lattice and the regions marked with 1(a)-1(f) and 2(a)-2(d) indicate the portions where ordered states in Fig.~\ref{fig:all-states} are stabilized. PS describes a phase separation region.  As in the previous figure, the spiral phase is stabilized for all values of $J_K$ at $n_e=0$ and $1$.}
		
	\end{figure}
	
	Next, we consider the strongly frustrated limit of the Shastry-Sutherland model where the strength of the interactions on the diagonal bonds is greater than the one on the axial bonds. The interest is driven by the fact that the system is in insulating state at $n_e=1/4$ and $3/4$ and when a non-coplanar order is stabilized at these number densities then there is a possibility that we can realize the quantized THE. Fig.~\ref{fig:nodm2} shows the phase diagram for $t'=2.2$, $J=0.1$, $J'=0.22$ and $\abs{\mathbf{D}}=\abs{\mathbf{D'}}=0$ on $80 \times 80$ lattice. The results for FM phase are the same as observed earlier. But now at $n_e\simeq1/2$ the ground state is not AFM but a coplanar flux state. For this configuration, the localized spins on the SSL diagonals are antiparallel aligned (singlet state). For most part of the phase diagram at small $J_K$ values the ground state is spiral state which with the increase of $J_K$ starts to become narrow in size. For large values of $J_K$, at $n_e=1/4$ and $3/4$ the phase diagram shows the PS regions. Similar to the previous case 3:1 collinear state is realized at in between values of number density. There are small region of all-out non-coplanar state in the phase diagram but not at $n_e=1/4$ and $3/4$.
	
	In the last part of our calculations, we \textit{turned on} the DM interaction in the limit we discussed in the last paragraph to realize the non-coplanar ground states. The results are shown in Fig.~\ref{fig:withdm2}, where phase diagram is plotted in the presence of DM interaction $D_\parallel=-0.1$, $D_\perp=0.15$ and $D'=-0.22$. Similar to what happened in the earlier case when we switched on the DM interaction here also the \textit{canted-Flux} becomes stabilized in the wider range of electron number density. In addition to $n_e=1/2$, this phase is stabilized at $n_e=1/4$ and $3/4$ for small and large values of $J_K$. Currently, we are using an unbiased finite temperature MC method to investigate the stability of this phase against thermal fluctuations and also exploring the magnetic and transport properties. The part in the phase diagram with all-out state becomes wider in the presence of DM interaction. FM phase is present at low and high values of $n_e$ similar to what we observed earlier. The coplanar flux state is totally replaced by \textit{canted-Flux} state. There are small patches of other coplanar states such as spiral, AFM and 3:1 collinear at some values of $n_e$.
	
	\section{Summary}\label{sec:summary}
	
	We have investigated the ground state phase diagram of Kondo lattice model on the SSL with classical localized spins. This study is relevant to exploring the novel magnetic and transport properties arising from the interplay between the charge-spin coupling and geometrical frustration in rare earth tetraborides. Using variational calculations, in addition to coplanar ordered states different non-collinear and non-coplanar arrangements of localized moments are realized over finite ranges of interaction parameters. Electron motion through such spin textures results in novel transport phenomena, such as the THE. The inclusion of DM interaction further stabilizes new \textit{canted-Flux} spin state.
	\vspace{0.8em}
	
   {\it \bf Acknowledgements:}
   \vspace{0.8em} 
	
   The work was partially supported by Grant No. MOE2014-T2-2-112 from the Ministry of Education, Singapore.

   \section{REFERENCES}

   \end{document}